\documentclass{jkas}

\def\year{2024} 
\def\volume{57} 
\def\issue{**} 
\def\beginpage{1} 
\def\received{September 21, 2024} 
\def\accepted{XX **, 2024} 
\def\published{XX **, 2024} 
\date{Received \received; Accepted \accepted; Published \published}

\def\cm3{~{\rm cm^{-3}}}

\usepackage{flushend} 
\usepackage{xcolor}

\title{Diffusive Shock Acceleration Efficiencies for Weak ICM Shocks in the Test Particle Regime}

\author{Hyesung Kang}{0000-0002-4674-5687}

\affil{Department of Earth Sciences, Pusan National University, Busan 46241, Korea; }

\def\corrauthor{H. Kang, \email{hskang@pusan.ac.kr}}

\def\runningauthor{Kang}
\def\runningtitle{Diffusive Shock Acceleration Efficiencies at Weak Shocks}

\def\keywords{acceleration of particles --- cosmic rays ---
galaxies: clusters: general --- shock waves}

\title{Diffusive Shock Acceleration Efficiencies for Weak ICM Shocks in the Test Particle Regime}

\def\abstracttext{
During the formation of large-scale structures in the universe, weak internal shocks are induced within the hot intracluster medium (ICM), while strong accretion shocks arise in the warm-hot intergalactic medium (WHIM) within filaments, and the warm-cold gas in voids surrounding galaxy clusters. These cosmological shocks are thought to accelerate cosmic ray (CR) protons and electrons via diffusive shock acceleration (DSA). Recent advances in particle-in-cell and hybrid simulations have provided deeper insights into the kinetic plasma processes that govern microinstabilities and particle acceleration in collisionless shocks in weakly magnetized astrophysical plasma.
In this study, we adopt a thermal-leakage type injection model and DSA power-law distribution functions in the test-particle regime. 
The CR proton spectrum directly connects to the Maxwellian distribution of protons at the injection momentum $p_{\rm inj} = Q_p p_{\rm th,p}$. On the other hand, the CR electron spectrum extends down to $p_{\rm min} = Q_e p_{\rm th,e}$ and is linked to the Maxwellian distribution of electrons. Here, $p_{\rm th,p}$ and $p_{\rm th,e}$, are the proton and electron thermal momenta, respectively. 
Moreover, we propose that the postshock gas temperature and the injection parameters, $Q_p$ and $Q_e$ are self-regulated to maintain the test-particle condition, as the thermal energy is gradually transferred to the CR energy.
Under these constraints, we estimate the self-regulated values of the temperature reduction factor, $R_T$, and the proton injection parameter, $Q_p$, along with the resulting CR efficiencies, $\eta_p$ and $\eta_e$.
We then provide analytical fitting functions for these parameters as functions of the shock Mach number, $M_s$.
These fitting formulas may serve as valuable tools for quantitatively assessing the impact of CR protons and electrons, as well as the resulting nonthermal emissions in galaxy clusters and cosmic filaments.
}

\begin{document}
\jkashead 

\section{Introduction\label{s1}}

\begin{figure*}[t]
\centering
\includegraphics[width=160mm]{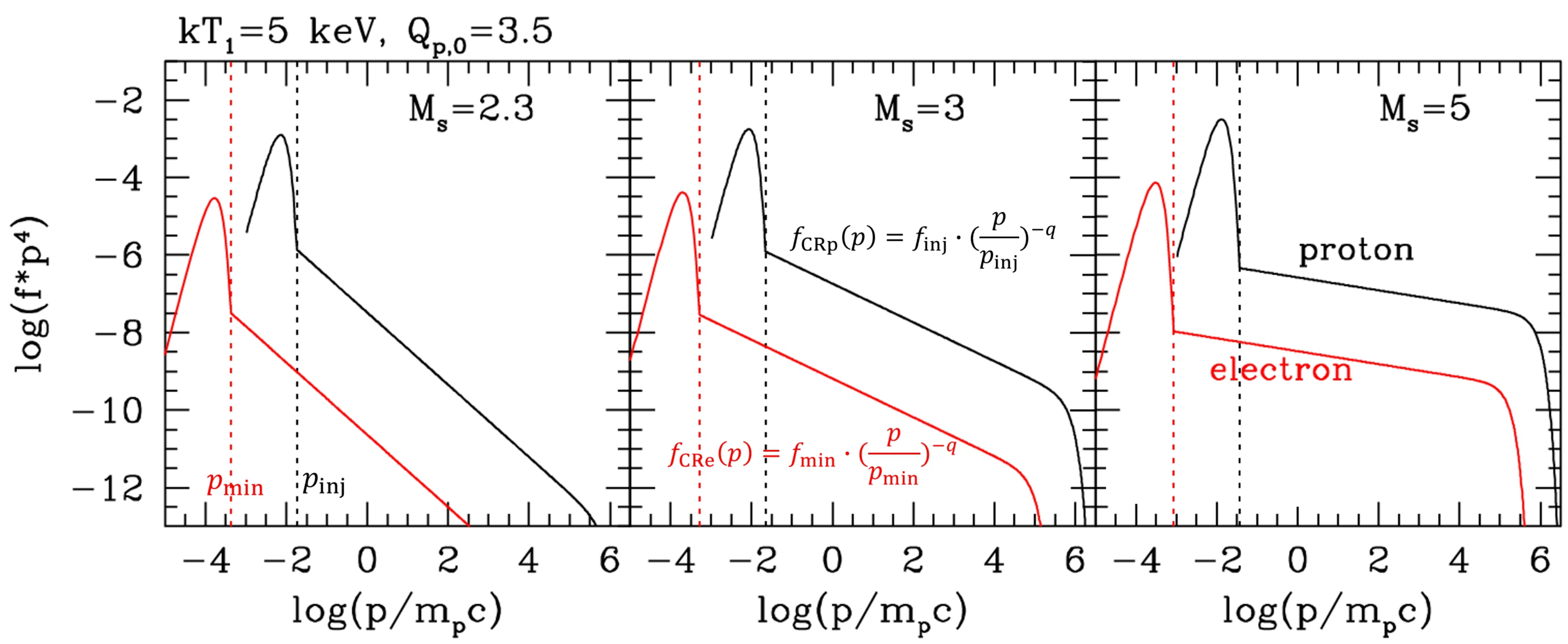}
\caption{Model power-law spectra, $f_{\rm CRp}(p)p^4$ (black) for protons and $f_{\rm CRe}(p)p^4$ (red) for electrons, are shown for shocks with $M_{\rm s} = 2.3$, 3, and 5 in the ICM, where the pre-shock temperature is $kT_1 = 5$~keV. For $f_{\rm CRp}(p)$, $p_{\rm max}/m_pc=10^6$ is adopted, while for $f_{\rm CRe}(p)$, $p_{\rm eq}$ is estimated using $B_2=1\mu{\rm G}$ and the redshift $z_r=0.2$. 
The red dotted line denotes $p_{\rm min} = Q_e p_{\rm th,e}$, above which suprathermal electrons are pre-accelerated in the shock transition zone. The black dotted line represents $p_{\rm inj} = Q_p p_{\rm th,p}$, above which suprathermal electrons and protons are accelerated by DSA by repeatedly crossing the shock transition layer. We assume injection parameters $Q_e \approx Q_p \approx 3.5$. The normalization factors, $f_{\rm min}$ and $f_{\rm inj}$, are given in Equations (\ref{fmin}) and (\ref{finj}), respectively. The momentum is expressed in units of $m_p c$. The absolute normalization of $f(p)$ is arbitrary. \label{f1}}
\end{figure*}

As the large-scale structure of the universe forms through hierarchical clustering of substructures, 
supersonic flow motions generate the so-called cosmological shocks in the cosmic web \citep[e.g.,][]{ryu2003,skillman2008,vazza2009,schaal2015,ha2018a}. 
These shocks can be broadly classified into two categories \citep[e.g.,][]{miniati2000,pfrommer2006, hoeft2008,hong2014,ha2023}: 
(1) {\it Internal shocks} occur in the hot intracluster medium (ICM), with $T \sim 10^7 - 10^8$ K, within the virial radius of galaxy clusters. They arise from ongoing mergers of substructures and turbulent flow motions. They have low sonic Mach numbers, $M_s \sim 2 - 4$, and a plasma beta, $\beta\equiv P_g/P_B\sim 50-100$, where $P_g$ and $P_B$ represent the gas and magnetic pressures, respectively.
(2) {\it External accretion shocks} form in the outer region of clusters, well outside of the virial radius.  They occur where the warm-hot intergalactic medium (WHIM) in filaments (with $T \sim 10^5 - 10^7$~K) and the warm-cold gas in void regions (with $T \sim 10^4$ K) accrete onto the cosmic web. Consequently, these shocks have the high sonic Mach numbers, $M_s \sim 10 - 10^2$ and their preshock gas is almost unmagnetized with $\beta\sim 10^2-10^3$(see Table 1 of \citet{ha2023}).

Similar to interplanetary shocks and supernova remnants, these shocks are expected to produce cosmic-ray (CR) protons and electrons through diffusive shock acceleration (DSA), resulting in a power-law distribution function, $f_{\rm DSA}\propto p^{-q}$, where $q=4M_s^2/( M_s^2-1)$ \citep[e.g.][]{drury1983,brunetti2014}.
The basic principle of DSA is well established: charged CR particles can gain energy by repeatedly crossing the shock front multiple times through scattering off underlying plasma or magnetohydrodynamic (MHD) waves \citep[e.g.,][]{bell1978, blandford1978, drury1983}.
For this process to be effective, however, suprathermal particles must first be pre-energized, kinetic plasma or MHD waves must be self-excited through various microinstabilities, or preexisting magnetic turbulence must be present in the background medium.

The injection and acceleration of CRs in collisionless shocks are governed by complex kinetic plasma processes that depend on various parameters, including the sonic Mach number $M_s$, the Alfv\'en Mach number $M_A$, the plasma beta $\beta$, and the obliquity angle $\theta_{\rm Bn}$ \citep[e.g.,][]{balogh2013,marcowith2016}.
In the weakly magnetized, high-$\beta$ ICM, low-Mach-number {\it internal} shocks are expected to predominantly accelerate protons in quasi-parallel configurations, where the obliquity angle $\theta_{\rm Bn}$, between the shock normal and the magnetic field, $\mathbf{B}_0$, is less than about $45^\circ$ \citep[e.g.,][]{ha2018b}. In contrast, electrons are preferentially accelerated in quasi-perpendicular configurations, where $\theta_{\rm Bn}$ exceeds $45^\circ$ \citep[e.g.,][]{guo2014,kang2019}. 
On the other hand, in strong, high-$\beta$, {\it accretion} shocks, electrons can be pre-energized and injected to DSA through stochastic Fermi II acceleration, due to the ion-Weibel filamentation instability in the shock transition \citep[e.g.,][]{weibel1959,medvedev2006,kato2010,bohdan2021,ha2023}.

The shock transition zone typically extends several times the postshock thermal proton gyroradius.
For both protons and electrons to cross the shock front and fully engage in the DSA process, they must be energized to reach the so-called injection momentum, $p_{\rm inj}$ \citep[e.g.,][]{caprioli2014,caprioli2015,park2015,ha2018b}. 
This initial stage of energizing CR particles, known as injection, remains one of the most significant unresolved challenges in DSA theory.

In this paper, the shock is characterized by the sonic Mach number $M_{\rm s}$, the shock speed $u_{\rm sh}$, the preshock gas density $\rho_1$, and temperature $T_1$, where the subscripts, $1$ and $2$, denote the preshock and postshock states, respectively.
Common physics symbols are used, such as $m_{\rm e}$ for the electron mass, $m_{\rm p}$ for the proton mass, $c$ for the speed of light, and $k$ for the Boltzmann constant.
For example, the thermal momenta for electrons and protons in the postshock region are $p_{\rm th,e}=(2m_e k T_{2})^{1/2}$ and $p_{\rm th,p}=(2m_p k T_{2})^{1/2}$, respectively.

Previous numerical studies using particle-in-cell (PIC) and hybrid simulations have shown that both protons and electrons can be extracted from the thermal pool and preaccelerated through scattering off kinetic waves that are self-excited in the shock transition zone and upstream \citep[e.g.,][]{caprioli2014, matsumoto2017, ha2018b, katou2019,kang2019, trotta2019,kobzar2021}. 
Similar to the classical thermal-leakage injection model \citep[e.g.,][]{malkov1998, gieseler2000,kang2002}, a small fraction of incoming thermal protons and electrons are reflected at the shock front by the electrostatic potential drop and/or the magnetic mirror force, resulting in {\it suprathermal tail} distributions in the energy spectra.
This preacceleration is expected to lead to their eventual injection into the DSA process \citep[see][and references therein]{kang2020,arbutina2021,ha2023}.

Based on the thermal-leakage injection model and previous plasma simulation studies, \citet{ryu2019} proposed a semi-analytic model for the momentum distribution function of CR protons, $f_{\rm CRp}(p)$, in which the postshock temperature, $T_2$, gradually decreases to maintain the test-particle condition as the thermal energy is progressively transferred to the CR energy (see their Figure 4).
In a subsequent study, \citet{kang2020} applied a similar DSA model for CR electrons, $f_{\rm CRe}(p)$.
In this framework, the DSA power-law spectra emerge from the respective Maxwellian distributions at $p_{\rm min}= Q_e p_{\rm th,e}$ for CR electrons and at $p_{\rm inj}= Q_p p_{\rm th,p}$ for CR protons (see Figure \ref{f1}). Here, the injection parameters are $Q_p\sim Q_e\sim 3.5$; $p_{\rm min}$ represents the smallest momentum of reflected electrons, and $p_{\rm inj}$ is the injection momentum above which CR particles fully participate in the DSA process.
In this study, we follow the same semi-analytic approach to calculate the self-regulated parameters, such as the temperature reduction factor $R_T$, the injection parameters $Q_p\approx Q_e$, and the resulting acceleration efficiencies for CR protons and electrons. We also provide analytic fitting functions for these parameters as functions of $M_s$.
These fitting functions are expected to offer practical prescriptions for CR acceleration in cosmological hydrodynamic or MHD simulations \citep[e.g.,][and reference therein]{vazza2016,wittor2017,pfrommer2017, winner2020,boss2023}. 

In the next section, we provide a detailed description of the semi-analytic models for $f_{\rm CRp}(p)$ and $f_{\rm CRe}(p)$
along the numerical approach used to self-adjust various parameters.
We then present the results for the self-regulated parameters and their corresponding fitting formulas as functions of the shock Mach number in Section \ref{s3}. A brief summary will be given in Section \ref{s4}.

\section{Model CR Spectra and Acceleration Efficiencies\label{s2}}

\subsection{Injection to DSA \label{s2.1}}

As mentioned in the introduction, the preacceleration of suprathermal particles is governed by a complex network of kinetic acceleration processes, including shock drift acceleration (SDA), stochastic SDA, Fermi I and II acceleration.
These mechanisms rely on the excitation of multi-scale plasma and MHD waves driven by various microinstabilities \citep[e.g.,][]{kim2021}. 
In turn, these processes depend on temperature anisotropies induced by shock-reflected particles \citep[e.g.,][and references therein]{ha2023}. As a result, the critical momenta, where the DSA power-law spectra connect to the respective Maxwellian distributions, can only be accurately determined through detailed PIC simulations that span a wide range of scales—from the electron skin depth to well beyond the gyroradii of suprathermal protons.

Since such extensive plasma simulations are currently not feasible, we rely on the results of previous numerical studies \citep[e.g.,][]{caprioli2014,ha2018b}, and assume that the injection momentum for CR protons can be parameterized as:
\begin{equation}
p_{\rm inj} = Q_p \cdot p_{\rm th,p},
\end{equation}
where the injection parameter $Q_{\rm p}\gtrsim 3.5$.

On the other hand, the electron injection requires a significantly more extended preacceleration, from $p_{\rm th,e}$ to $p_{\rm inj} = Q_p (m_p/m_e)p_{\rm th,e}$.
Following \citet{kang2020}, we assume that the power-law spectrum of suprathermal electrons begins approximately at the lowest momentum of the {\it reflected electrons}, such that
\begin{equation}
p_{\rm min}= Q_e \cdot p_{\rm th,e}.
\end{equation}
For simplicity, we assume that the electron injection parameter is similar to that of protons, with $Q_e\approx Q_p$ \citep[e.g.,][]{kang2019,arbutina2021}.
The relative relation between $p_{\rm min}$ and $p_{\rm inj}$ is illustrated by the vertical red and black dashed lines in Figure \ref{f1}.

\subsection{Analytic DSA Power-law Spectrum \label{s2.2}}

\begin{figure}[t]
\centering
\includegraphics[width=80mm]{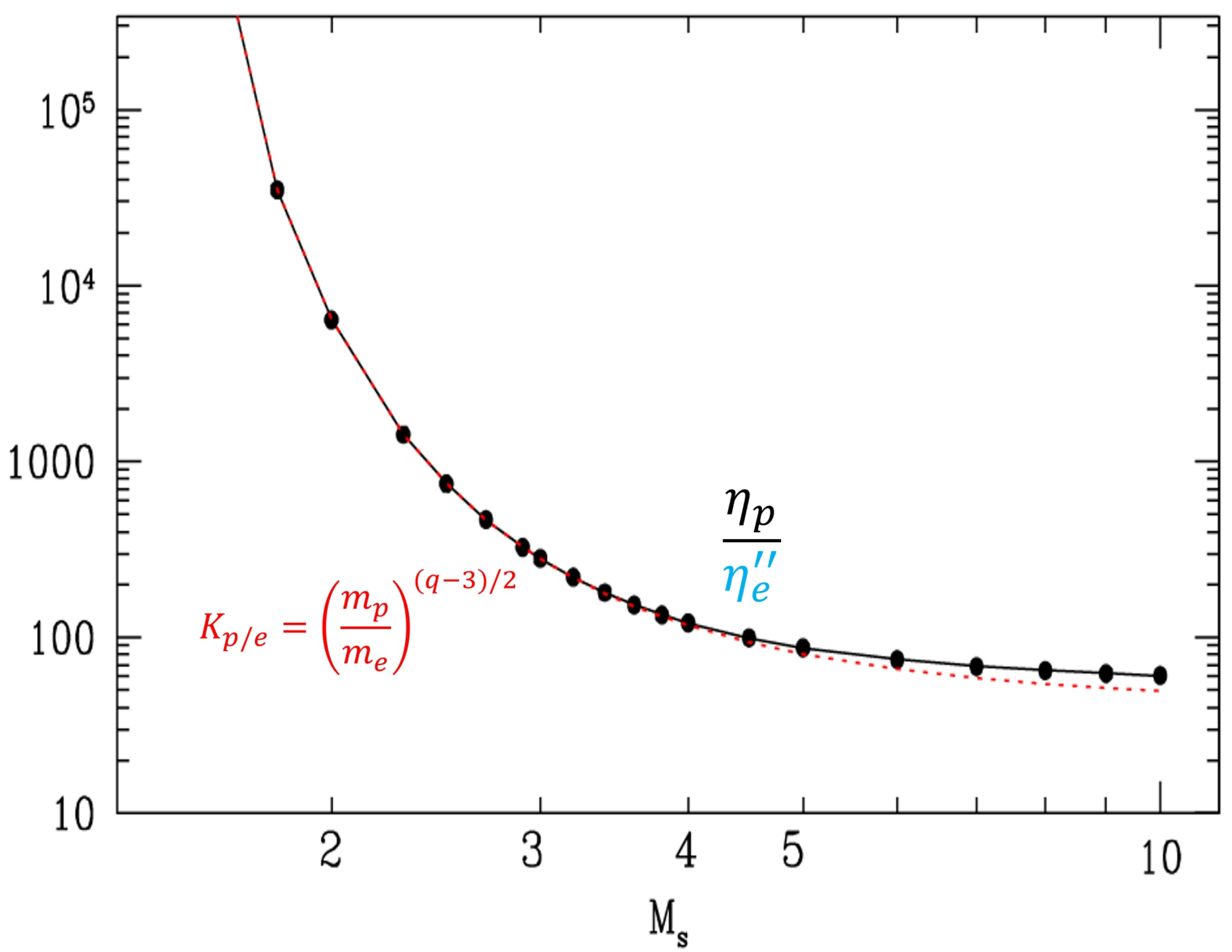}
\caption{ 
CR proton-to-electron number ratio, $K_{p/e}=(m_p/m_e)^{(q-3)/2}$, is shown with the red dotted line.
For comparison, the ratio of $\eta_p/\eta_e^{\prime\prime}$ is displayed for the case of $kT_1 = 5$~keV and $Q_{p,0} = 3.5$, as shown in Fig \ref{f3}.
\label{f2}}
\end{figure}

The spectrum of CR protons is assumed to follow a power law for $p\ge p_{\rm inj}$:
\begin{equation}
f_{\rm CRp}(p) \approx f_{\rm inj}\cdot \left(p \over p_{\rm inj} \right) ^{-q} \exp\left(-{p^2 \over p_{\rm max}^2} \right) .
\label{fpinj}
\end{equation}
The normalization factor, $f_{\rm inj}$, is specified by the Maxwellian distribution at $p_{\rm inj}$:
\begin{equation}
f_{\rm inj} = {n_{\rm p,2} \over \pi^{1.5}} p_{\rm th,p}^{-3} \exp(-Q_p^2),
\label{finj}
\end{equation}
where $n_{\rm p,2}$ is the postshock proton number density.
By default, the initial value is set as  $Q_{\rm p,0}=3.5$, while $T_2$ and $Q_{\rm p}$ are self-consistently adjusted to ensure that the CR proton energy remains within the test-particle regime as $p_{\rm max}$ increases, i.e., $\eta_p\le0.03$.
 
Similarly, the power-law spectrum of CR electrons is represented for $p\ge p_{\rm min}$:
\begin{equation}
f_{\rm CRe}(p) \approx f_{\rm min}\cdot \left(p \over p_{\rm min} \right) ^{-q} \exp\left(-{p^2 \over p_{\rm eq}^2} \right) .
\label{feinj}
\end{equation}
The normalization factor, $f_{\rm min}$, is again specified by the Maxwellian distribution at $p_{\rm min}$:
\begin{equation}
f_{\rm min} = {n_{\rm e,2} \over \pi^{1.5}} p_{\rm th,e}^{-3} \exp(-Q_e^2),
\label{fmin}
\end{equation}
where $n_{\rm e,2}$ is the postshock electron number density and $Q_e=Q_p$.
The cutoff momentum, $p_{\rm eq}$, can be derived from the equilibrium condition that the DSA momentum gains per cycle
are equal to the synchrotron and inverse Compton losses per cycle. These losses depend on the magnetic field strength, $B_2$, and the energy density of the cosmic background radiation \citep[e.g.,][]{kang11}.
As illustrative examples, we show the thermal Maxwellian distributions and the test-particle power-law spectra, $f_{\rm CRp}(p)$ and $f_{\rm CRe}(p)$, for shocks with $M_s=2.3-5$ and $kT_1=5$~keV in Figure \ref{f1}.

\begin{figure*}[t]
\centering
\includegraphics[width=160mm]{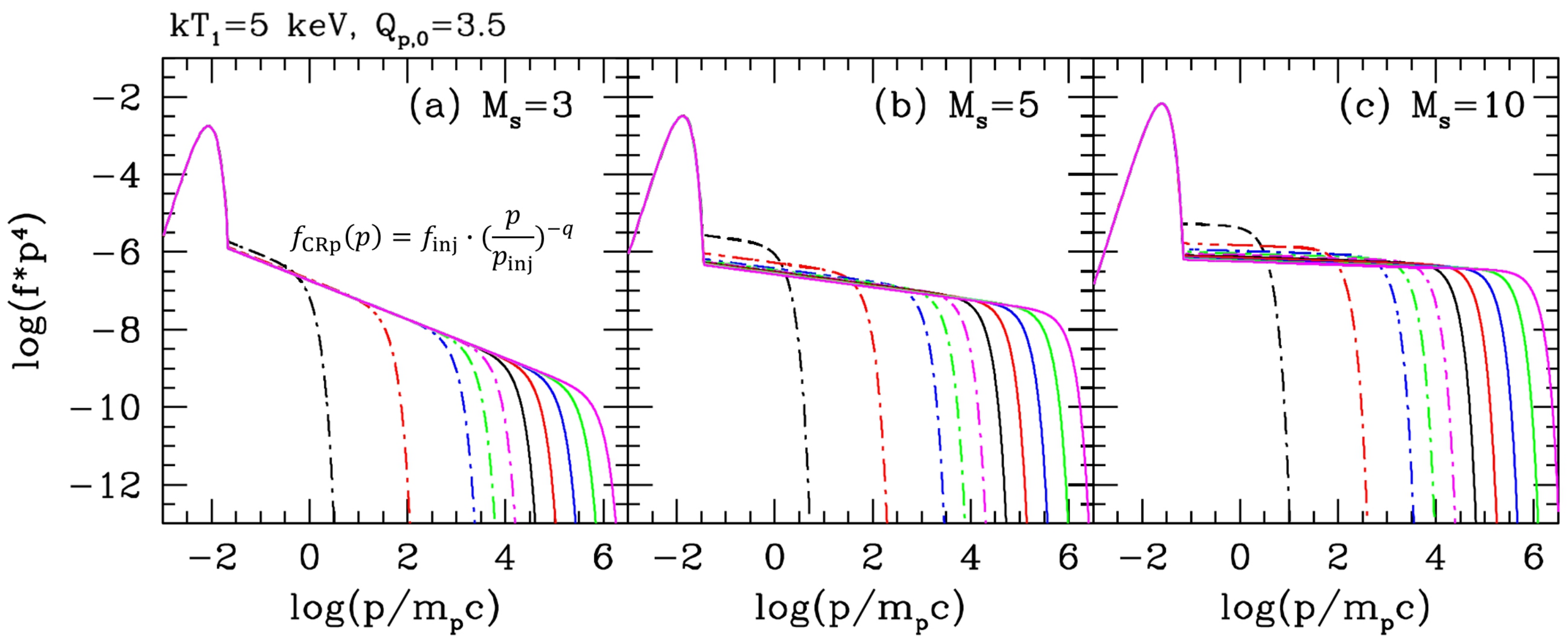}
\caption{Model power-law spectra for protons, $f_{\rm CRp}(p)p^4$, for $M_{\rm s} = 3$, 5, and 10 shocks in the ICM at $kT_1 = 5$~keV. As the maximum momentum, $p_{\rm max}$, gradually increases, the injection parameter $Q_p$ is adjusted to increasingly larger values, resulting in a lower value of the normalization factor $f_{\rm inj}$. The initial value for the injection parameter is $Q_{p,0} = 3.5$.
\label{f3}}
\end{figure*}

\subsection{DSA Injection Number Fractions\label{s2.3}}

For the model spectra in Equations (\ref{fpinj}) and (\ref{feinj}), the CR injection fractions by number, $\xi_p$ and $\xi_e$, are solely determined by the injection parameters, $Q_p$ and $Q_e$, and the power-law index, $q$:
\begin{equation}
\xi_{p,e}\equiv \frac{n_{\rm CRp,e}}{n_2} \approx \frac{4}{\sqrt{\pi}}Q_{p,e}^3 {\rm exp}(-Q_{p,e}^2)\frac{1}{q-3}. 
\label{xipe}
\end{equation}
This expression assumes that both $p_{\rm max}\!\gg\! 1$ and $p_{\rm eq}\!\gg\!1$. 
Under these conditions, if $Q_p \approx Q_e$ is assumed, 
 the number of CR particles injected to DSA is approximately the same for protons and electrons, i.e., $N_{\rm CRp}\approx N_{\rm CRe}$.
With these DSA model spectra, the CR proton-to-electron number ratio, $K_{p/e}$, is given by the ratio of $f_{\rm CRp}$ to $f_{\rm CRe}$ at  $p=p_{\rm inj}$:
\begin{equation}
K_{p/e} \approx \frac{f_{\rm CRp}(p_{\rm inj})}{f_{\rm CRe}(p_{\rm inj})} = \left({ p_{\rm th,p} \over {p_{\rm th,e}}}\right)^{q - 3} = \left({m_{p} \over m_{e}}\right)^{(q - 3)/2},
\label{Kpe}
\end{equation}
which depends only the mass ratio, $m_p/m_e$ and the power-law index $q$.
The red line in Figure \ref{f2} illustrates this relationship as a function of $M_s$.
For strong shocks, where $q \approx 4$, the value of $K_{p/e}$ is approximately 43, given $m_p/m_e = 1836$. 
As $q$ increases, $K_{p/e}$ also increases, meaning that lower Mach number shocks yield larger ratios.

\begin{figure*}[t]
\centering
\includegraphics[width=160mm]{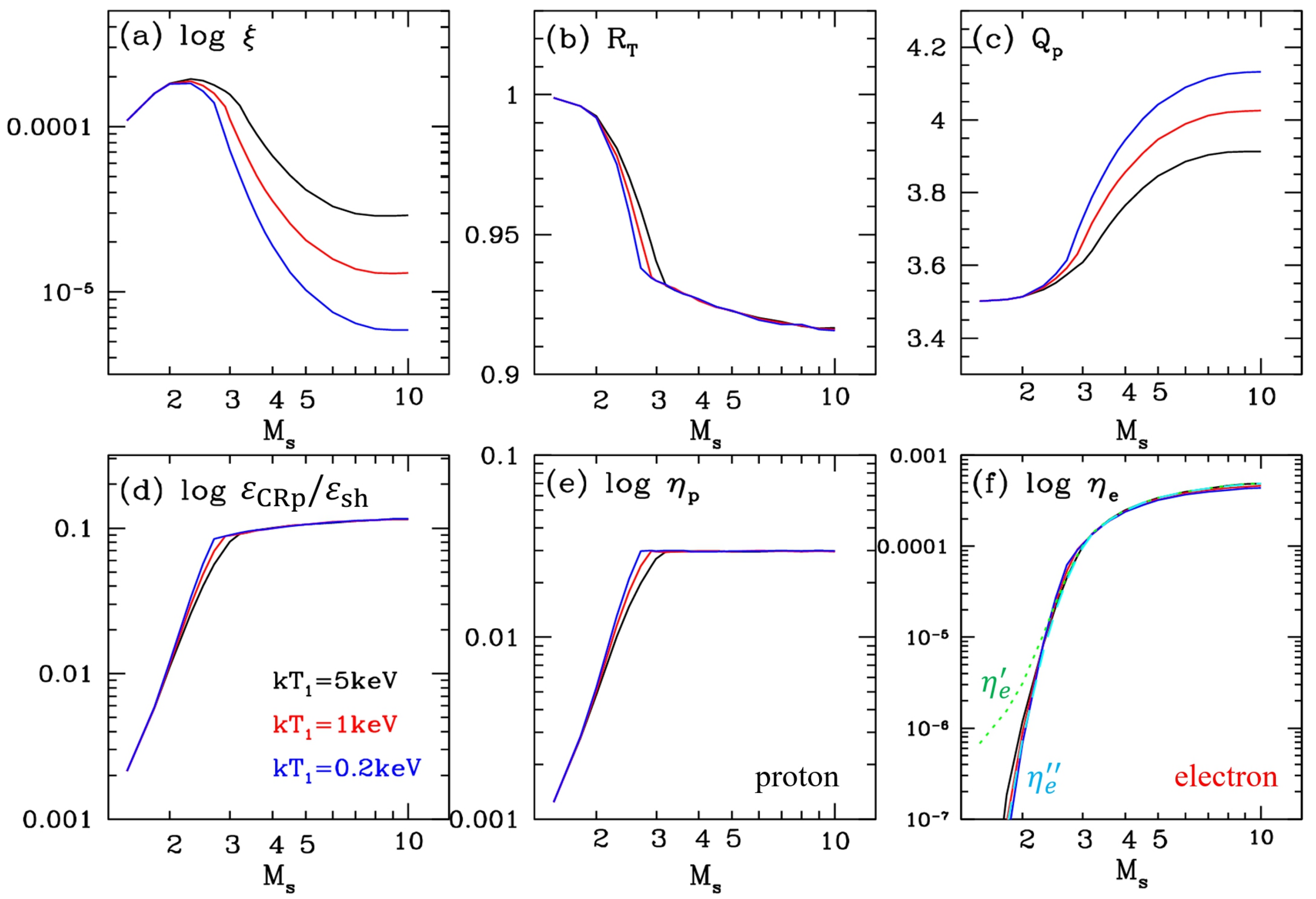}
\caption{The injection number fraction, $\xi_p = \xi_e$, the temperature reduction factor, $R_{\rm T}$, the self-adjusted injection parameter for protons, $Q_p$, the postshock CR energy fraction, $\mathcal{E}_{\rm CRp}/\mathcal{E}_{\rm sh}$, the CR proton acceleration efficiency, $\eta_p$, and the CR electron acceleration efficiency, $\eta_e$, are plotted as functions of $M_{\rm s}$.	
The results are shown for preshock temperatures of $k T_1= 0.2$ (blue), 1 (red), and 5~keV (black).
The initial value for the proton injection parameter is $Q_{p,0} = 3.5$, which increases gradually to maintain $\eta_p\le 0.03$ as the maximum proton momentum increases to $p_{\rm max} = 10^7~m_p c$.
For the CR electron acceleration efficiency, the lowest momentum is assumed to be $p_0 = 10~m_e c$ for the fiducial case (black, red, and blue solid lines).
In panel (f), for comparison, the green dotted line depicts $\eta_e^{\prime}$ with $p_0 = p_{\rm min}$, while the cyan dashed line shows $\eta_e^{\prime\prime}$ with $p_0 = p_{\rm inj}$.
Note that $p_{\rm min} < 10~m_e c < p_{\rm inj}$. \label{f4}}
\end{figure*}

\subsection{Enforcement of Test-Particle Condition \label{s2.4}}

As in previous studies, we define the DSA acceleration efficiency as the ratio of the post-shock CR energy flux to the shock kinetic energy flux \citep{ryu2019}:

\begin{eqnarray}
\eta_p\equiv \frac{\mathcal{E}_{\rm CRp} u_2}{0.5\rho_1 u_{\rm sh}^3}=\frac{1}{r}\frac{\mathcal{E}_{\rm CRp}}{\mathcal{E}_{\rm sh}},\\ \nonumber
\eta_e\equiv \frac{\mathcal{E}_{\rm CRe} u_2}{0.5\rho_1 u_{\rm sh}^3}=\frac{1}{r}\frac{\mathcal{E}_{\rm CRe}}{\mathcal{E}_{\rm sh}}.
\label{eta}
\end{eqnarray}
Here, $r$ is the density compression ratio across the shock front, $\mathcal{E}_{\rm CRp}$ and $\mathcal{E}_{\rm CRe}$ are the energy densities of CR protons and electrons, respectively, at the shock location, and $\mathcal{E}_{\rm sh}=(1/2)n_1 m_p u_{\rm sh}^2$ is the shock kinetic energy density. Note that $\mathcal{E}_{\rm CRe}\ll \mathcal{E}_{\rm CRp}$ due to the small electron to proton mass ratio, $m_e/m_p$.
We assume that the CR proton acceleration occurs within the test-particle regime, imposing the condition $\eta_p\le 0.03$, which is approximately equivalent to $\mathcal{E}_{\rm CRp} \lesssim 0.1 \mathcal{E}_{\rm sh}$. 
This upper limit on $\eta_p$ is somewhat arbitrary but is motivated by the expectation that CR acceleration in ICM shocks occurs within the test-particle regime, where the DSA power-law spectrum remains valid.
In particular, the non-detection of gamma-ray emission from galaxy clusters provides a reasonable constraint on $\eta_p$ \citep[e.g.,][]{ha2020,wittor2020}. 

In this approach, we follow the numerical method outlined in \citet{ryu2019}, where the post-shock gas temperature decreases self-consistently as shock energy is transferred to CR protons.
First, we calculate the energy density of CR proton using the following equation:
\begin{equation}
\mathcal{E}_{\rm CRp} = 4 \pi c \int_{p_0}^{p_1} (\sqrt{p^2+ (m_{p}c)^2}-m_{p}c) f_{\rm CRp}(p) p^2 dp.
\label{ECR}
\end{equation}
where $p_0$ and $p_1$ represent the lower and upper bounds of the CR proton momentum distribution, respectively.
In the cases of the power-law spectrum, $f_{\rm CRp}\propto p^{-q}$,
the energy density increases approximately as $\mathcal{E}_{\rm CRp}\propto \ln p_1$ for strong shocks with $q=4$. 
For weaker shocks with $q>4$, $\mathcal{E}_{\rm CRp}$ depends on both $p_0$ and $q$.
We set $p_0=p_{\rm inj}$ and $p_1=p_{\rm max}$, gradually increasing $p_{\rm max}$ up to $10^7 m_pc$.

As $p_{\rm max}$ and $\mathcal{E}_{\rm CRp}$ increase incrementally, the post-shock temperature gradually decreases from $T_{2,0}$, which is derived from the purely gasdynamic condition with $\mathcal{E}_{\rm CRp}=0$, to $T_2= R_T \cdot T_{2,0}$.
To account for the transfer of thermal energy $\mathcal{E}_{\rm th}$ to CRp energy $\mathcal{E}_{\rm CRp}$,
the temperature reduction factor is estimated as follow:
\begin{equation}
 R_T=\frac{ \mathcal{E}_{\rm th}(T_{2,0})-\mathcal{E}_{\rm CRp}}{\mathcal{E}_{\rm th}(T_{2,0})}\le 1. 
\end{equation}
The proton injection parameter then increases approximately as $Q_p\approx Q_{p,0}/\sqrt{R_T}$.
By reducing $T_2$ and increasing $Q_p$, the proton acceleration efficiency remains $\eta_p\le 0.03$ (or $\mathcal{E}_{\rm CRp} \lesssim 0.1 \mathcal{E}_{\rm sh}$).
Once $R_T$ and $Q_p$ are fixed, $\mathcal{E}_{\rm CRp}$ is calculated using Equation (\ref{ECR}).

For the energy density of CR electrons, $\mathcal{E}_{\rm CRe}$, a formula similar to Equation (\ref{ECR}) is used, with $m_p$ replaced by $m_e$ and appropriate values chosen for $p_0$ and $p_1$.
For the upper bound, we set $p_1=p_{\rm eq}$, where the equilibrium momentum $p_{\rm eq}$ is estimated based on typical values for ICM shocks, such as the postshock magnetic field strength, $B_2=1~\mu$G, and the redshift, $z_r=0.2$.
However, the results are not sensitive to these choices, as $p_{\rm eq}/m_ec\gtrsim 10^5  \gg 1$.
For the lower bound, we consider the three options:
(1) $\eta_e$ calculated with $p_0=10m_ec$, which serves as the fiducial case.
(2) $\eta_e^{\prime}$ calculated with $p_0=p_{\rm min}$, which includes the suprathermal electrons.
(3) $\eta_e^{\prime\prime}$ calculated with $p_0=p_{\rm inj}$, which is consistent with the definition of CR particles capable of crossing  the shock transition zone.

\section{Results \label{s3}}

According to Equation (\ref{Kpe}), $K_{\rm p/e}\approx \eta_p/\eta_e^{\prime\prime}$, provided that $p_{\rm max}\!\gg\!1$ and $p_{\rm eq}\!\gg1\!$, $\eta_e^{\prime\prime}$ is calculated with the lower bound, $p_0=p_{\rm inj}$. 
This ratio is represented the black circles in Figure \ref{f2}.
We note that the black circles are slightly higher than the red line for high $M_s$, because $p_{\rm max}$ is higher than $p_{\rm eq}$.

Figure \ref{f3} illustrates the self-adjustment of $f_{\rm CRp}(p)$ for shocks with $M_s=3-10$ and $kT_1=5$~keV as $p_{\rm max}$ increases from  $10 m_pc$ to $10^6 m_pc$. 
The initial value of the injection parameter was set to $Q_{\rm p,0}=3.5$.
For the $M_s=3$ shock, the self-adjustment is minimal, as shown in Panel (a).
In contrast, for shocks with $M_s=5$ and $10$, $Q_p$ gradually increases from $3.5$ to $3.9$, while $T_2$ decreases slightly, with $R_T\approx 0.92$, as $p_{\rm max}$ increases.
These changes result in a slightly larger $p_{\rm inj}$ and a lower $f_{\rm inj}$, as shown in Panels (b)-(c).

\begin{figure}[t]
\centering
\includegraphics[width=90mm]{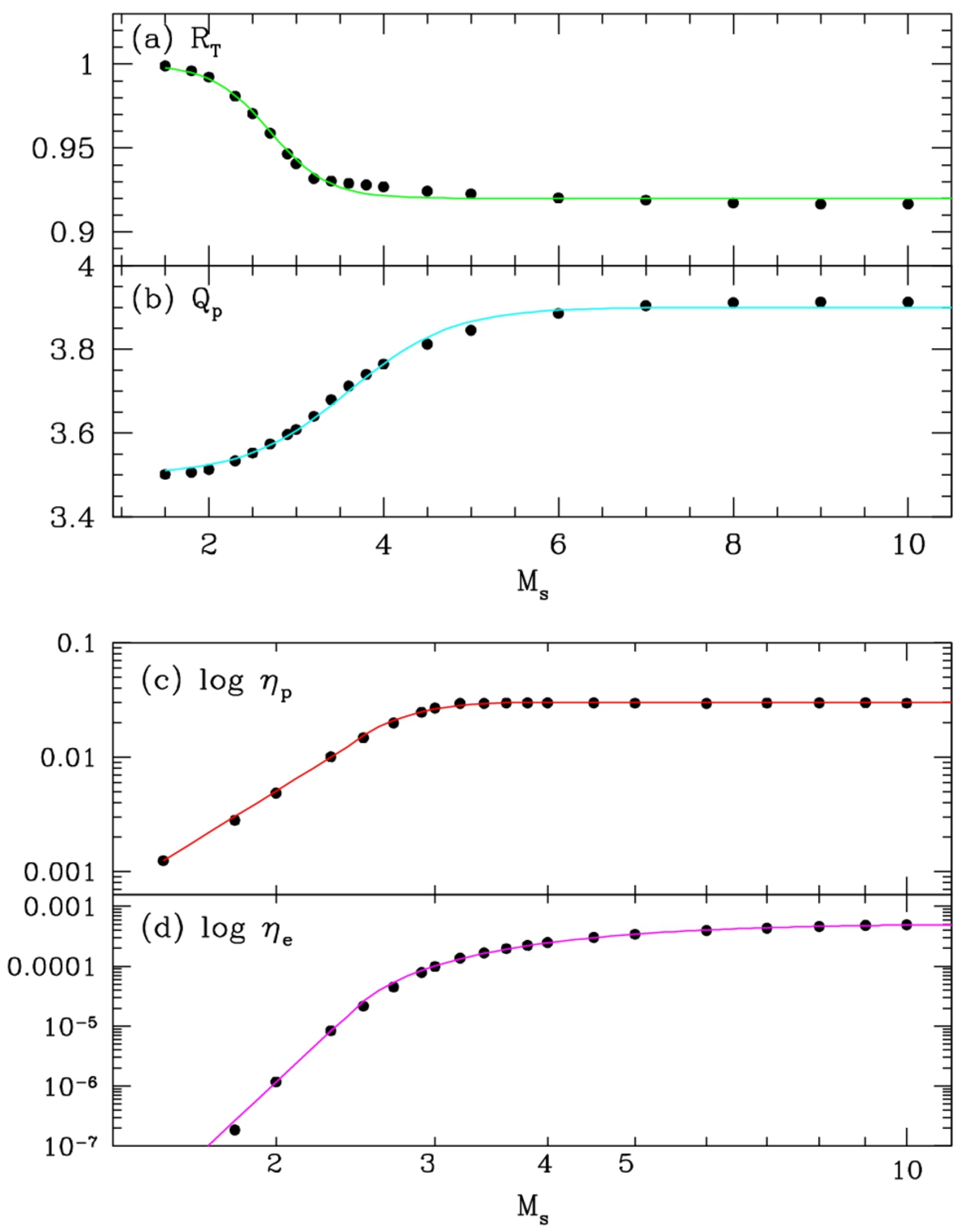}
\caption{Fitting functions for $R_{\rm T}$ (green), $Q_p$ (cyan), $\eta_p$ (red), and $\eta_e$ (magenta) in Equations (\ref{RT})-(\ref{etae}) are displayed for the case of $kT_1 = 5$~keV and $Q_{p,0} = 3.5$. For comparison, the black circles represent the numerical data, as shown in Figure \ref{f4}.
\label{f5}}
\end{figure}

Figure \ref{f4} presents the results for three sets of our model calculations for the preshock temperatuer $kT_1=0.2$~keV (blue), 1~keV (red), and 5~keV (black), to explore the dependence on the preshock temperature.
Panels (d) and (e) show that the CR proton acceleration efficiency is constrained within $\eta_p\le0.03$, while $\mathcal{E}_{\rm CRp}\lesssim 0.1 \mathcal{E}_{\rm sh}$.
This is achieved by shifting $R_T$ to a smaller value and $Q_p$ to a larger value, as shown in Panels (b) and (c).
By doing so, the injection number fraction, $\xi_p$, decreases for higher $M_s$.
The degree of this self-adjustment for higher $M_s$ shocks is more pronounced at lower $T_1$.
We take the models with $kT_1=5$~keV as the fiducial cases, since the main focus is on weak ICM shocks.

However, Figures \ref{f4}(e)-(f) show that the acceleration efficiency for CR protons, $\eta_p$, depends only weakly on the preshock temperature, while the acceleration efficiency for CR electrons, $\eta_e$, is almost independent of $T_1$.
In Panel (f), the three cases for $kT_1=5$~keV—$\eta_e$ with $p_0=10m_ec$ (black solid line), $\eta_e^{\prime}$ with $p_0=p_{\rm min}$ (green dotted line), and $\eta_e^{\prime\prime}$ with $p_0=p_{\rm inj}$ (cyan dashed line)—illustrate the dependence on the lower bound of the energy-density integral. 
For $M_s< 3.0$, the relation $\eta_e^{\prime}<\eta_e<\eta_e^{\prime\prime}$ holds, while for $M_s> 3.0$ the three values converge. This is because, for shocks with $M_s> 3.0$, the CR energy spectrum flattens and becomes increasingly dominated by high-energy particles.

We adopt functional forms to fit the results of our modeling as follows.  
In Figures \ref{f5}(a)-(b), the black circles represent the results for our fiducial model with $kT_1=5$~keV, while the green and cyan lines show their fitting forms for $R_T$ and $Q_p$, respectively:
\begin{equation}
 R_T(M_s) \approx  1.0-\frac{0.08}{1+ \exp[-3(M_s-2.7)]},
\label{RT}
\end{equation}

\begin{equation}
 Q_p(M_s) \approx 3.5+\frac{0.4}{1+ \exp[-1.7(M_s-3,6)]}.   
\label{Qp}
\end{equation}
Our model spectra, given in Equations (\ref{fpinj})-(\ref{fmin}), can be specified using these fitting forms for $R_T(M_s)$ and $Q_p(M_s)$, provided that the shock parameters, including $M_s$, $\rho_1$, and $T_1$, are given.

Figures \ref{f5}(c)-(d) show the numerical results and their corresponding fitting forms for $\eta_p$ and $\eta_e$:

 \begin{equation}
 \eta_p(M_s) \approx \left \{\begin{array}{lll}
      0.01\cdot(M_s/2.3)^{4.88},\quad M_s< 2.3;\\
      \Sigma_{n=0}^4 a_n (M_s-1)^n/M_s^4,\quad  2.3\le M_s< 3.5;\\
      0.03, \quad  3.5\le M_s\le 10,
    \end{array}
  \right.
\label{etap}
\end{equation}
where the fitting coefficients are $a_0= 4.03$, $a_1= -8.80$, $a_2= 6.11$, $a_3= -1.44$, and $a_4=0.171$.
            
\begin{equation}
 \eta_e(M_s) \approx \left \{\begin{array}{lll}
      8.38\times 10^{-6}\cdot(M_s/2.3)^{14.1},\quad M_s< 2.3;\\
      \Sigma_{n=0}^4 b_n (M_s-1)^n/M_s^4,\quad  2.3\le M_s< 10,
    \end{array}
  \right.
\label{etae}
\end{equation}
where the fitting coefficients are $b_0= -2.85\times10^{-3}$, $b_1= 1.12\times10^{-2}$, $b_2= -1.32\times10^{-2}$, $b_3= 4.37\times10^{-3}$, and $b_4=4.05\times10^{-4}$.

These fitting functions can be used to estimate $\mathcal{E}_{\rm CRp}$ and $\mathcal{E}_{\rm CRe}$ from $\mathcal{E}_{\rm sh}$ using Equation (9).
In particular, the fitting functions for $\eta_p$ can be compared with Equations (41)-(42) and Figure 1 presented in \citet{boss2023}.

\section{Summary \label{s4}}

In low-Mach-number cosmological shocks that occur in weakly magnetized ICM plasma, the pressure contribution from CR particles is likely to be dynamically insignificant, suggesting that DSA operates in the test-particle regime.
The non-detection of gamma-ray emission from galaxy clusters, in particular, provides a reasonable constraint on the CR proton acceleration efficiency \citep[e.g.,][]{ha2020,wittor2020}. 
Under these conditions, the CR proton and electron distribution functions follow a DSA power-law form of $p^{-q}$, where the spectral index $q$ is determined solely by the shock Mach number, $M_s$.

In this study, we assume that the CR proton spectrum, $f_{\rm CRp}(p)$, connects directly to the proton Maxwellian distribution at the injection momentum $p_{\rm inj} = Q_p p_{\rm th,p}$, while the electron spectrum, $f_{\rm CRe}(p)$, extends down to $p_{\rm min} = Q_e p_{\rm th,e}$, linking to the electron Maxwellian distribution (see Figure \ref{f1}).
The precise values of the injection parameters, $Q_p$ and $Q_e$, remain uncertain due to the complexity of plasma kinetic processes involved in the preacceleration or injection of suprathermal particles into the DSA mechanism, as well as the practical challenges of performing PIC simulations over the broad range of scales required—from electron plasma to ion gyro scales. We propose that $Q_p$ and $Q_e$ are self-regulated as the post-shock thermal energy density gradually decreases, with CR energy increasing as $p_{\rm max}$ increases (see Figure \ref{f3}).

Using the power-law spectra defined in Equations (\ref{fpinj})-(\ref{feinj}) and calculating the CR energy density via Equation (\ref{ECR}), we estimate the self-regulated values of $R_T$ and $Q_p$, along with the resulting CR efficiencies, $\eta_p$ and $\eta_e$, as functions of $M_s$ for shocks with preshock temperatures of $kT_1 = 0.2$, 1, and 5 keV (see Figure \ref{f4}). 
We impose the test-particle condition, $\eta_p \leq 0.03$, or equivalently, $\mathcal{E}_{\rm CRp} \lesssim 0.1 \mathcal{E}_{\rm sh}$, to self-consistently decrease $R_T$ while increasing $Q_p$.
This upper limit on $\eta_p$ is based on the expectation that CR acceleration in ICM shocks occurs in the test-particle regime, where the DSA power-law spectrum remains valid.
While $Q_p(M_s)$ shows a weak dependence on $T_1$, the other three quantities exhibit only marginal variations.

Finally, we provide analytical fitting functions for $R_T$, $Q_p$, $\eta_p$, and $\eta_e$ for the case of $kT_1 = 5$ keV in Equations (\ref{RT})-(\ref{etae}) (see Figure \ref{f5}). The CR spectra, $f_{\rm CRp}(p)$ and $f_{\rm CRe}(p)$, can be reconstructed using these fitting formulas for $R_T$ and $Q_p$, given shock parameters such as $M_s$, $\rho_1$, and $T_1$. Alternatively, the CR energy densities, $\mathcal{E}_{\rm CRp}$ and $\mathcal{E}_{\rm CRe}$, can be estimated from the shock kinetic energy density, $\mathcal{E}_{\rm sh} = (1/2) \rho_1 u_{\rm sh}^2$, using the fitting formulas for $\eta_p$ and $\eta_e$. Thus, these numerical fitting formulas are valuable for studies aimed at quantitatively evaluating the influence of CRs and the associated nonthermal emissions from galaxy clusters and filaments \citep[e.g.][]{ha2023}.

One thing that should be noted is that it is still not clear if injection to DSA through scattering off self-generated kinetic waves is possible for subcritical shocks with $M_s< 2.3$ \citep[e.g.][]{ha2018b,kang2019}.
It was demonstrated that the rippling of the shock surface, excited by Alfv\'en Ion Cyclotron (AIC) instability, could induce multi-scale fluctuations only at supercritical shocks with $M_s\gtrsim 2.3$, leading to the pre-acceleration of electrons beyond $p_{\rm inj}$ \citep{trotta2019,ha2021,ha2022,kobzar2021,boula2024}.
However, this issue needs further investigation for high-$\beta$ shocks in the presence of preexisting turbulence \citep[e.g.,][]{guo2015}.


\acknowledgments
This work was supported by the National Research Foundation (NRF) of Korea through grant 2023R1A2C1003131.





\end{document}